\documentclass[aps,prd,twocolumn,superscriptaddress]{revtex4}

\usepackage{bm}
\usepackage{amsfonts,amssymb}
\usepackage{amsmath}
\usepackage{graphicx,graphics}
\usepackage{enumerate}
\usepackage{subfigure}
\usepackage{appendix}
\usepackage{siunitx}
\usepackage{mathrsfs}  
\usepackage{hyperref}

\newcommand{\Acte}{k}

\begin{document}

\title{Analogue model for anti-de Sitter as a description of point sources in fluids}

\author{Ricardo A. Mosna}
\email{mosna@ime.unicamp.br}
\affiliation{Departamento de Matem\'atica Aplicada, Universidade Estadual de Campinas, 13083-859, Campinas, SP, Brazil}
\author{Jo\~ao Paulo M. Pitelli}
\email{pitelli@ime.unicamp.br}
\affiliation{Departamento de Matem\'atica Aplicada, Universidade Estadual de Campinas, 13083-859, Campinas, SP, Brazil}
\author{Maur\'icio Richartz}
\email{mauricio.richartz@ufabc.edu.br}
\affiliation{Centro de Matem\'atica, Computa\c{c}\~ao e Cogni\c{c}\~ao, UFABC, 09210-170 Santo Andr\'e, SP, Brazil}

\begin{abstract}
We introduce an analogue model for a nonglobally hyperbolic spacetime in terms of a two-dimensional fluid. This is done by considering the propagation of sound waves in a radial flow with constant velocity. We show that the equation of motion satisfied by sound waves is the wave equation on $AdS_2\times S^1$. Since this spacetime is not globally hyperbolic, the dynamics of the Klein-Gordon field is not well defined until boundary conditions at the spatial boundary of $AdS_2$ are prescribed. On the analogue model end, those extra boundary conditions provide an effective description of the point source at $r=0$. For waves with circular symmetry, we relate the different physical evolutions to the phase difference between ingoing and outgoing scattered waves. We also show that the fluid configuration can be stable or unstable depending on the chosen boundary condition.
\end{abstract}

\maketitle

\section{Introduction}
\label{sec:intro}

The flow of an inviscid fluid is governed by three equations: the continuity equation, Euler's equation and an equation of state. In the case of an irrotational and barotropic fluid, the resulting linear perturbation $\delta \Psi$ of the velocity potential of the flow satisfies a massless Klein-Gordon equation on a curved background,
\begin{equation} 
\label{KG} 
\frac{1}{\sqrt{-g}}\partial_{\mu}(\sqrt{-g}g^{\mu \nu} \partial_{\nu}\delta \Psi)=0.
\end{equation}
The effective metric $g_{\mu \nu}$, with determinant $g=\det{(g_{\mu \nu})}$, is related to the background density and velocity in an algebraic fashion~\cite{unruh} (see Ref.~\cite{barcelo-liberati-visser} for a detailed review about this topic).  

Analogue gravity models of this kind are often used to probe kinematical aspects of general relativity and quantum field theory on curved spacetimes, such as cosmological particle production~\cite{jain-weinfurtner, schutzhold-uhlmann, schutzhold-unruh}, superradiance~\cite{basak-majumdar,basak-majumdar-2,berti-cardoso-lemos,richartz-prain-liberati-weinfurtner,cardoso-coutant-richartz-weinfurtner} and Hawking radiation~\cite{brout-massar, barcelo-liberati-visser-2, giovanazzi, schutzhold-unruh-3}. In particular, the transition between sub- and supercritical fluid flows can be set up in a laboratory, generating a dumb hole, the analogue of a black hole~\cite{unruh,visser}, which allows the classical analogue of the Hawking radiation to be tested experimentally \cite{rousseaux-mathis-maissa,weinfurtner-tedford-penrice,parentani}. The quantum version of this effect has also been studied experimentally~\cite{belgiorno,unruh_comment,belgiorno_reply}, with very exciting results~\cite{steinhauer1,steinhauer2}.  

In this paper, we present an analogue model of a nonglobally hyperbolic spacetime without an event horizon.  Our focus is on $AdS_2\times S^1$, the metric of which is given by 
\begin{equation}
ds^2=\frac{-d\tau^2+dr^2}{r^2}+\gamma^2 d\theta^2,
\label{ads2s1}
\end{equation}
where $\tau$ is the time coordinate, $(r,\theta)$ are spatial coordinates and $0<\gamma<1$. Notice that for $\theta=\textrm{const}$, this metric represents the Poincar\'e patch of an $AdS_2$ spacetime, which is nonglobally hyperbolic and, therefore, requires a boundary condition at the spacelike infinity for one to solve initial value problems. This is because there is a boundary at the spacelike infinity of anti de-Sitter (AdS) so that information can effectively flow in from infinity. In the coordinates just given, the spatial infinity is located at $r=0$, and, in the analogue model introduced here, this corresponds to an external source (or sink) of fluid in an inviscid flow in $\mathbb{R}^2$. The lack of global hyperbolicity of $AdS_2$ is then associated with the need to model how the wave interacts with the point source/sink at $r=0$. As we briefly review below, the specification of a boundary condition at the spacelike infinity is related to the prescription of a self-adjoint extension of the wave operator in AdS. In this way, the analogue model considered here provides a physical interpretation of this abstract mathematical procedure in terms of an effective description of the point source. 

This paper is organized as follows. In Sec.~\ref{sec:model}, we discuss our model, which consists of a two-dimensional fluid with constant radial velocity coming out or falling into a source/sink. Sound waves propagating in this flow obey the massless Klein-Gordon equation in  $AdS_2\times S^1$. In Sec.~\ref{sec:BC}, we briefly review how the lack of global hyperbolicity of the anti-de Sitter space calls for a boundary condition at $r=0$, which corresponds to a choice of a self-adjoint extension of the (spatial part of) the wave operator. This allows us to study, in Sec.~\ref{sec:scattering}, the  scattering properties of sound waves in this fluid/spacetime. We show that the boundary condition is encoded in the phase difference between incoming and outgoing waves and that the stability of the configuration depends on the chosen boundary condition. Finally, the last section contains our conclusions.

\section{Two-dimensional fluid with radial flow}
\label{sec:model}

Consider an inviscid fluid with density $\rho=\rho(r)$ and velocity $\vec{v}=v(r)\hat{r}$ in a stationary, two-dimensional flow. If the flow is irrotational, the velocity $\vec{v}$ can be written in terms of a scalar potential $\Psi$, $\vec{v} = -\nabla \Psi$. As discussed in the Introduction, linear perturbations $\delta \Psi$ of the scalar potential obey Eq.~(\ref{KG}). The associated $2+1$ acoustic metric is~\cite{unruh, barcelo-liberati-visser} 
\begin{equation}
g_{\mu\nu}=\left(\frac{\rho}{c}\right)^2\left[\begin{array}{cc}
-(c^2-v^2)& -\vec{v}^{T}\\
-\vec{v}& I_{2\times 2}
\end{array}\right],
\end{equation}
where $I_{2\times2}$ is the identity matrix. We assume here a constant speed of sound $c$ which amounts to an equation of state of the form $p=\textrm{constant}\times\rho$.

Due to the two-dimensional continuity equation in a stationary regime, 
\begin{equation}
\nabla\cdot\left(\rho \vec{v}\right)=\frac{1}{r}\frac{\partial}{\partial r}\left[r \rho(r) v(r)\right]=0,
\end{equation} 
the density is related to the flow velocity by $\rho(r)=\Acte/(rv(r))$, where $\Acte$ is a constant. Although this flow is somehow exotic, we show next that it works perfectly as a toy model for nonglobally hyperbolic spacetimes in analogue gravity. 
This leads to the line element
\begin{equation*}
\begin{aligned}
ds^2&=\left(\frac{\Acte}{crv}\right)^2\left[-(c^2-v^2)dt^2-2vdtdr+dr^2+r^2d\theta^2\right]\\
&=\left(\frac{\Acte}{crv}\right)^2\bigg[-(c^2-v^2)\left(dt+\frac{v}{c^2-v^2}dr\right)^2\\&+\left(1+\frac{v^2}{c^2-v^2}\right)dr^2+r^2d\theta^2\bigg]\\
&=\left(\frac{\Acte}{crv}\right)^2\left[-(c^2-v^2)d\tau^2+\left(\frac{c^2}{c^2-v^2}\right)dr^2+r^2d\theta^2\right],
\end{aligned}
\end{equation*}
with $d\tau=dt+v \, dr/(c^2-v^2)$. Since $v=v(r)$, no integrability issues arise here.

We now assume that the fluid has a constant subsonic velocity $\vec{v}=\alpha c \hat{r}$, with $-1<\alpha<1$. This fluid configuration represents a radially flowing fluid which comes out from a source (if $\alpha>0$) or falls into a sink (if $\alpha <0$) located at the origin. 
This yields
\begin{equation}
ds^2=\frac{\Acte^2}{\alpha^2c^4r^2}\left[-c^2(1-\alpha^2)d\tau^2+\frac{dr^2}{1-\alpha^2}+r^2d\theta^2\right],
\label{conformal cone}
\end{equation}
where $\tau=t+\frac{\alpha}{c(1-\alpha^2)}r$.

If we further define $\bar{\tau}=c (1-\alpha^2) \tau$ and $\lambda=\sqrt{1-\alpha^2}$ then the metric can be recast as 
\begin{equation}
ds^2=\text{const} \times \left( \frac{-d\bar{\tau}^2+dr^2}{r^2} +\lambda^2 d\theta^2\right),
\label{AdS cone}
\end{equation}
which represents the $AdS_2\times S^1$ spacetime. Notice that, for $\theta=\textrm{constant}$, Eq.~(\ref{AdS cone}) represents a Poincar\'e patch of the $AdS_2$ spacetime. 

\section{Boundary conditions}
\label{sec:BC}

The propagation of waves in nonglobally hyperbolic spacetimes was first considered by Wald \cite{wald} and then by Ishibashi and Wald \cite{ishibashi-wald-1,ishibashi-wald-2}. They showed that the spatial portion of the wave equation is generally not self-adjoint in these cases. It is only by picking an arbitrary positive definite self-adjoint extension of this operator that a fully deterministic dynamical evolution can be prescribed.

Consider a static spacetime with a timelike Killing field $\xi^{\mu}$ and a static spacelike slice $\Sigma$. (Notice that $\Sigma$ is not a Cauchy surface since the spacetime is nonglobally hyperbolic). The wave equation for a minimally coupled massless scalar field $\nabla^\mu\nabla_\mu\psi=0$ can then be split into the form
\begin{equation}
\frac{\partial^2\psi}{\partial \tau^2}=VD^{i}(VD_{i}\psi)\equiv -A \psi,
\label{split}
\end{equation}
where $\tau$ denotes the Killing parameter, $V^2=-\xi^{\mu}\xi_\mu$, and $D_{i}$ is the covariant derivative on $\Sigma$. Let the domain of the operator $A$ be given by the set of smooth functions with compact support, i.e., $\mathcal{D}(A)=C_0^{\infty}(\Sigma)$. The operator $A$ is a positive symmetric operator in the Hilbert space $\mathcal{H}$ of the square integrable functions on $\Sigma$ with measure $V^{-1}d\Sigma$. Hence, it has at least one self-adjoint extension $\bar{A}$ (the closure of $A$) \cite{reed-simon}. If there is only one such extension, the evolution is uniquely determined by the background spacetime and is given by
\begin{equation}
\psi(\tau)=\cos{\left(\bar{A}^{1/2}\tau\right)}\psi(0)+\bar{A}^{-1/2}\sin{\left(\bar{A}^{1/2}\tau\right)}\dot{\psi}(0).
\end{equation}
However, if there is an infinite number of self-adjoint extensions $A_E$, with $E$ being a parameter, the evolution is ambiguous, and one extra condition---which specifies which extension should be picked---must be given. To each such extension $A_{E}$, there corresponds a different physical evolution given by
\begin{equation}
\psi_E(\tau)=\cos{\left(A_E^{1/2}\tau\right)}\psi(0)+A_E^{-1/2}\sin{\left(A_E^{1/2}\tau\right)}\dot{\psi}(0).
 \end{equation}

In order to determine the boundary conditions, we start by splitting the acoustic metric (\ref{conformal cone}) into
\begin{equation}
ds^2=-V(r)^2d\tau^2+h_{ij}dx^{i}dx^{j},
\end{equation}
where $V(r)=\frac{\Acte(1-\alpha^2)^{1/2}}{\alpha c r}$ and $h_{ij}$, the metric on the spatial slice $\Sigma$, is given by
\begin{equation}
h_{ij}=\left(\begin{array}{cc}
\frac{\Acte^2}{\alpha^2c^4r^2(1-\alpha^2)} & 0\\
0& \frac{\Acte^2}{\alpha^2c^4}
\end{array}\right).
\end{equation}
A simple but tedious calculation following Eq.~\eqref{split} then shows that sound waves satisfy
\begin{equation}
\frac{\partial^2\Psi}{\partial \tau^2}=c^2(1-\alpha^2)^2\frac{\partial^2\Psi}{\partial r^2}+\frac{c^2(1-\alpha^2)}{r^2}\frac{\partial^2\Psi}{\partial \theta^2},
\label{wave equation}
\end{equation}
with measure $d\mu\equiv V^{-1}d\Sigma=\frac{\Acte}{\alpha c^3(1-\alpha^2)}drd\theta$ (for simplicity, we have dropped the $\delta$ in $\delta \Psi$).

If we consider sound waves with circular symmetry, i.e., $\Psi\equiv \Psi(r,\tau)$, then Eq.~(\ref{wave equation}) becomes
\begin{equation}
\frac{\partial^2\Psi}{\partial \tau^2}=c^2(1-\alpha^2)^2\frac{\partial^2\Psi}{\partial r^2},
\label{simpler equation}
\end{equation}
with $r>0$. The operator $A=-\frac{d^2}{dr^2}$ with initial domain $C_0^{\infty}(0,\infty)$ is symmetric. Its Hilbert adjoint operator is given by $A^{\ast}=-\frac{d^2}{dr^2}$ with domain $L^{2}(0,\infty)$. Since these domains are different, in order to obtain a unitary evolution, one needs to enlarge the domain of $A$ (so that the domain of $\mathcal{D}(A^{\ast})$ is reduced) until both domains become identical. This amounts to specifying the right boundary conditions at $r=0$. 

The general theory is explained in detail in~Ref. \cite{reed-simon}, but the central objects of this procedure are the kernels $N_{\mp}$ of the operators $A^{\ast}\pm i$. If these spaces have the same dimension $n$, then there is a family of self-adjoint extensions of $A$ parametrized by the isometries from $N_{+}$ to $N_{-}$, i.e., by $n\times n$ unitary matrices. As a result, each such isometry $U$ defines a new (enlarged) domain given by \cite{reed-simon}
\begin{equation}\begin{aligned}
\mathcal{D}(A_U)=&\big\{\varphi_0+\varphi_++U\varphi_+|\varphi_0\in \mathcal{D}(A),\\& \varphi_+\in\text{Ker}(A^{\ast}-i)\big\}.
\end{aligned}
\end{equation}
Going back to our problem, for which $A=-\frac{d^2}{dr^2}$, the solutions of $(A^{\ast}\pm i)\varphi_{\mp}=0$ are given by
\begin{equation}\begin{aligned}
&\varphi_{+}=C_1e^{-\frac{1-i}{\sqrt{2}}r}+C_2e^{\frac{1-i}{\sqrt{2}}r},\\
&\varphi_{-}=C_3e^{-\frac{1+i}{\sqrt{2}}r}+C_4e^{\frac{1+i}{\sqrt{2}}r}.
\end{aligned}
\end{equation}
The normalizable solutions are the ones which decay exponentially, and this yields $C_2=C_4=0$. Therefore, we end up with $\varphi_{+}=e^{-\frac{1-i}{\sqrt{2}}r}$ and $\varphi_-=e^{-\frac{1+i}{\sqrt{2}}r}$. In this way, we have $n=1$ so that the isometries from $\text{Ker}(A^{\ast}-i)$ to $\text{Ker}(A^{\ast}+i)$ are parametrized by $1\times1$ unitary matrices, i.e., by phases $U_{\theta}=e^{i\theta}$. As a result, a function $\varphi(r)$ belongs to $\mathcal{D}(A_U)$ if it has the form
\begin{equation}
\varphi(r)=\varphi_0(r)+e^{-\frac{1-i}{\sqrt{2}}r}+e^{i\theta}e^{-\frac{1+i}{\sqrt{2}}r}.
\end{equation}
Since $\varphi_0(0)=\varphi_0'(0)=0$, we have
\begin{equation}\begin{aligned}
\frac{\varphi'(0)}{\varphi(0)}&=\frac{1}{\sqrt{2}}\left[-1-i\frac{1-e^{i \theta}}{1+e^{i\theta}}\right]\\&=-\frac{1}{\sqrt{2}}\left(1+\tan(\theta/2)\right)\equiv -\frac{1}{\beta}.
\end{aligned}\end{equation}
Therefore, the self-adjoint extensions $A_\beta$ of the operator $A$ are related to the boundary conditions by
\begin{equation}
\varphi(0)+\beta\varphi'(0)=0.
\label{boundary condition}
\end{equation}
Note that the parameter $\beta$ has the dimension of length~\cite{dimension}.

This is a good place to note that, since the radial velocity of the fluid is constant, there is no predefined (natural) length scale associated to the propagating sound waves. It is the self-adjoint extension which sets the length scale $\beta$ of the problem.

If we relax our assumptions and consider, instead of circular symmetry, sound waves with an $m$-fold rotational symmetry, i.e., $\Psi=\Psi(r,\tau)e^{im\theta}$, the wave equation becomes 
\begin{equation}
\frac{\partial^2\Psi}{\partial \tau^2}=c^2(1-\alpha^2)^2\frac{\partial^2\Psi}{\partial r^2}-\frac{c^2(1-\alpha^2)m^2}{r^2}\Psi.
\end{equation}
This represents an inverse square potential problem with $A\propto-\frac{d^2}{dr^2}+\frac{m^2}{(1-\alpha^2)r^2}$. For $m\geq 1$, this problem does not need an extra boundary condition since $m^2/(1-\alpha^2)\geq 3/4$ (for details, see Ref. \cite{gitman-tyutin-voronov}). Accordingly, only  the case of circular symmetry ($m=0$) will be of interest to us.

\section{Scattering and Stability}
\label{sec:scattering}

We now investigate how the scattering of an incident circular wave is affected by the choice of the self-adjoint extension. In other words, we impose the boundary condition \eqref{boundary condition} and study how the associated solution of the wave equation depends on the parameter $\beta$. We also analyze the stability of the fluid configuration by looking for solutions with imaginary frequencies.

\subsection{Oscillating modes}
\label{sec:oscillating}

The general solution of Eq.~(\ref{simpler equation}) corresponding to the scattering of a circularly symmetric sound wave is generated by superposition of harmonics,
\begin{equation}
\Psi^{(\omega)}=\left[\eta e^{-i\frac{ \omega }{c(1-\alpha^2)}r}+\xi e^{i\frac{ \omega }{c(1-\alpha^2)}r}\right]e^{-i \omega \tau},
\label{incoming-outgoing}
\end{equation}
where $\eta$ and $\xi$ are ($\omega$-dependent) constants. It follows from Eq.~(\ref{boundary condition}) that
\begin{equation}
\eta\left(1-i\frac{\beta\omega}{c(1-\alpha^2)}\right)+\xi\left(1+i\frac{\beta\omega}{c(1-\alpha^2)}\right)=0.
\label{bc}
\end{equation}

\begin{figure}[h!]
    \includegraphics[width=0.48\textwidth]{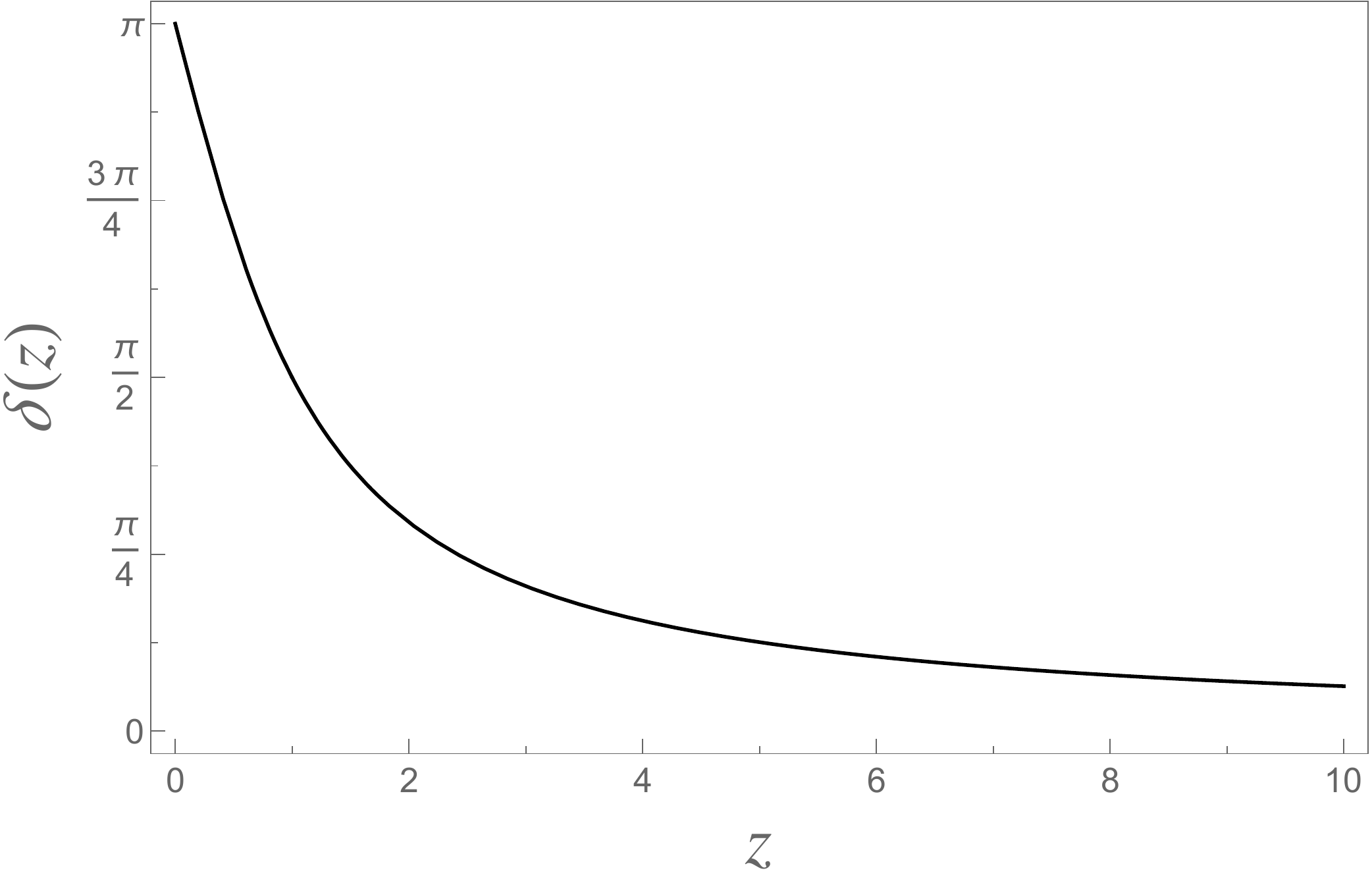}
		\caption{The phase difference between the incoming and the outgoing waves as a function of the dimensionless parameter $z=\frac{\beta\omega}{c(1-\alpha^2)}$.}
		\label{fig1}
\end{figure}

For real $\omega$, the phase difference $\delta(\omega)$ between the incident and reflected waves is then given by
\begin{equation} \label{relations}
e^{i \delta}=\frac{\xi}{\eta}=\frac{\frac{i\beta \omega}{c(1-\alpha^2)}-1}{\frac{i\beta \omega}{c(1-\alpha^2)}+1}
\end{equation}
(notice that $|\xi/\eta|=1$, as expected). We note that this phase difference is a function of the dimensionless parameter
$$
z\equiv\frac{\beta\omega}{c(1-\alpha^2)},
$$
so that it is $\beta$ which sets the scale for $\omega$ (or $\omega/c(1-\alpha^2)$). The phase difference profile is shown in Fig.~\ref{fig1}. Notice that different choices for $\beta$ correspond to rescalings of $\omega$. In particular, for $\beta=0$ (Dirichlet boundary condition), the phase difference is constant and equal to $\pi$, while for $\beta=\pm\infty$ (Neumann boundary condition), $\delta\equiv 0$. We will return to this point in Sec.~\ref{sec:conclusion}.

In terms of the original variables of the problem, Eq.~(\ref{incoming-outgoing}) becomes
\begin{equation}
\Psi= \left[\eta e^{-i\frac{ \omega }{c(1-\alpha)}r}+\xi e^{i\frac{ \omega }{c(1+\alpha)}r}\right]e^{-i \omega t},
\label{incoming-outgoing2}
\end{equation}
and we clearly see that the incident and reflected waves correspond to sound waves with velocities $c\pm v$, which makes physical sense. More specifically, for a source (sink), i.e., $\alpha>0$ ($\alpha<0$), the first term in Eq.~(\ref{incoming-outgoing2}) represents an incoming circular wave with velocity $c-v$ ($c+v$) traveling against (along with) the fluid flow, while the second term represents an outgoing wave with velocity $c+v$ ($c-v$) traveling along with (against) the fluid flow. 

It is worth noting that quantum field theory is well established in globally hyperbolic spacetimes, wherein the Friedrichs extension of the Helmholtz operator is unique. The Friedrichs extension is also the usual choice for AdS, which corresponds to $\beta=0$ in our notation, and this leads to the calculation of several physical quantities (such as two-point functions~\cite{allen}) and physical predictions (see, for instance, Ref. \cite{burges} for a derivation of the trace anomaly in AdS). However, these quantities will generally be affected by the boundary conditions at spatial infinity, i.e., by the choice of the self-adjoint extension~\cite{hadamard}. Therefore, identifying the particular boundary condition of a given system is a decisive step for studying its physics. In the analogue model considered here, the boundary condition can be determined in terms of a measurable, observable quantity, namely the phase difference between the incident and the reflected waves.

\subsection{Damped modes}
\label{sec:damped}

Now, we consider the case of damped or exponentially growing modes, which correspond to imaginary values of $\omega$ in Eq.~(\ref{incoming-outgoing}).

It is easy to see that for $\beta<0$ there are no solutions of this kind that are finite as $r\to\infty$ and satisfy the boundary condition (\ref{boundary condition}). As a result, the oscillatory modes discussed above exhaust the spectrum of solutions. The same happens for $\beta=0,\pm \infty$. Therefore, in these cases, the solutions are mode stable.

On the other hand, for $0<\beta<+\infty$, there are two solutions of this kind, which are given by
\begin{align}
&\Psi=e^{-(1+\alpha) r/\beta}e^{-c(1-\alpha^2) t/\beta},\label{1}\\
&\Psi=e^{-(1-\alpha) r/\beta}e^{c(1-\alpha^2) t/\beta},\label{2}
\end{align}
with $\omega=-ic(1-\alpha^2)/\beta$ and $\omega=ic(1-\alpha^2)/\beta$, respectively. Equation (\ref{1}) represents a mode which is damped in both $r$ and $t$ and is consequently stable. However, Eq.~(\ref{2}) represents a mode with finite energy but that grows exponentially in time. Therefore, for $0<\beta<+\infty$, the fluid configuration is unstable under linear perturbations. 

\section{Conclusions}
\label{sec:conclusion}

To summarize, we have proposed an analogue model of a nonglobally hyperbolic spacetime. Our construction uses, once again, the analogy between sound waves traveling on a background fluid flow and scalar fields propagating on a curved spacetime. We found that, on a two-dimensional fluid with constant radial velocity, sound waves satisfy the wave equation on an $AdS_2\times S^1$ spacetime. The lack of global hyperbolicity of $AdS_2$ introduces an ambiguity in the evolution of circularly symmetric sound waves in such a background. This is intimately related to the theory of self-adjoint extensions of symmetric operators. We showed that, in order to uniquely specify the wave evolution, one needs to impose an extra boundary condition at the source (or sink) of the fluid at the origin, which corresponds to the spatial infinity of $AdS_2$. 

A detailed study of scattering of circular waves showed that choosing this extra boundary condition is tantamount to specifying the phase difference between the incident and the reflected waves (as seen in Fig.~\ref{fig1}). This provides a physical interpretation of the mathematical formalism of self-adjoint extensions of operators in terms of an effective description of source/sink at $r=0$. Not all boundary conditions are physical, though. It is only for $\beta<0$ and $\beta=0,\pm\infty$ that the system is stable (to first order). 

Our analysis was based on the propagation of sound waves in a fluid, but almost any other analogue model of gravity could be used. For example, if surface waves propagating on an open channel are considered~{\cite{unruh_ralf}, one simply has to replace the density of the fluid $\rho(r)$ by the depth of the fluid $h(r)$ and the speed of sound $c$ by the speed of surface waves $\sqrt{a_g h(r)}$, where $a_g$ is the gravitational acceleration.

From a strictly mathematical point of view, the model studied here is equivalent to a one-dimensional semi-infinite string with its end attached to a spring with elastic constant proportional to $-\beta$ \cite{Graff} (the limiting cases of $\beta=0$ and $\beta=\pm\infty$ correspond to a string with a fixed or free end, respectively). The unstable case of Sec.~\ref{sec:damped} thus corresponds to an inverted harmonic oscillator, which makes it clear why the system generically runs out of control for $\beta>0$: the sound waves may continuously gain energy from (or lose energy to) the source/sink.  On the other hand, under this mathematical equivalence, for $\beta<0$, the string and the spring continuously exchange energy in a stable way. These stable configurations correspond to {\em positive} extensions of the operator $A$ in the sense of Ref. \cite{ishibashi-wald-2}, and it is precisely for this kind of extensions that the energy
\begin{equation}
E=\frac{1}{2}\int{\left(c^2(1-\alpha^2)|\partial_r \Psi|^2+|\partial_\tau \Psi|^2\right)d\mu}
\end{equation} 
is conserved. It is not difficult to show that, for the simple system considered here, this energy is essentially the same as the energy of sound waves in a fluid~\cite{Stone}. This wraps up our discussion in a nice way.

It is worth mentioning that the self-adjoint extension method used here is a valuable tool in solving the wave equation in classically singular spacetimes (which is not the case of anti-de Sitter). Our results can be easily extended to this case. In this context, a classically singular spacetime is said to be quantum mechanically nonsingular when there is only one self-adjoint extension of the spatial part of the wave operator (so that there is no ambiguity in the evolution of the quantum field). When there is an infinite number of such extensions, the spacetime is said to be quantum mechanically singular~\cite{horowitz}. The present work can be taken as a starting point to the study of analogous models of quantum singularities. 

We finally note that the supersonic counterpart of the model presented here can be shown to be associated with the de Sitter spacetime. This leads to interesting physical and mathematical consequences which will be the subject of a subsequent paper.

\acknowledgments

The authors are grateful to D. Q. Aruquipa, A. Saa and V. B. Silveira for stimulating discussions and to an anonymous referee for feedback. The authors acknowledge support from FAPESP Grant No. 2013/09357-9. J. P. M. P. also acknowledges support from FAPESP Grant No. 2016/07057-6 and FAEPEX Grant No. 2693/16.


\begin{thebibliography}{99}
\bibitem{dimension}  The procedure that led to Eq.~(\ref{boundary condition}) requires one to solve the equation $(A^{\ast}\pm ik_0^2)\varphi=0$, where $k_{0}$ is an arbitrary positive constant with the dimension of $1/(\textrm{length})$. Following the usual practice, we took $k_{0}=1$ in the that derivation.
\end{thebibliography}
\end{document}